\documentstyle[preprint,aps]{revtex}
\begin{document}
\draft
\title{Analytic expressions for the single particle energies with a
quadrupole-quadrupole interaction and the relation to Elliott's SU(3) model}
\author{E. Moya de Guerra$^1$, P. Sarriguren$^1$ and L. Zamick$^2$}
\address{$^1$ Instituto de Estructura de la Materia, Consejo Superior de\\
Investigaciones Cient\'\i ficas, Serrano 123, E-28006 Madrid, Spain\\
$^2$ Department of Physics and Astronomy, Rutgers University, Piscataway,\\
New Jersey 08855, USA}
\date{\today }
\maketitle

\begin{abstract}
We present an analytical proof and a generalization of the
Fayache-Sharon-Zamick relation between single particle energy splittings and
the SU(3) limit in Elliott's model.
\end{abstract}

\pacs{21.60.-n,21.10.-k}

\section{Introduction}

In a recent publication~\cite{FSZ}, Fayache, Sharon and Zamick compared the
collective model result for a rotational band to the 0d-1s splitting
obtained with the OXBASH program \cite{oxbash}. For the shell model
calculations done with OXBASH they use a Hamiltonian consisting of a
spherical harmonic oscillator and a residual interaction which is the
isoscalar quadrupole-quadrupole (Q-Q) in coordinate space only, i.e., they
omit the $Q\cdot Q$ term in momentum space and allow for interactions
between shells with $\Delta N=2$. From these shell model calculations the
authors of Ref. \cite{FSZ} found that one third of the 0d-1s single particle
(s.p.) splitting (6$\overline{\chi }$) comes from the interaction of the
valence particle with the core while two thirds (12$\overline{\chi }$) come
from the diagonal $Q\cdot Q$ interaction. The 0d-1s s.p. energy-splitting (18%
$\overline{\chi }$) is the same as the energy splitting between L=2 and L=0
rotational bands obtained with Elliott's SU(3) model in the s-d shell~\cite
{Elliott}. A similar numerical relation was found in the f-p shell~\cite{FSZ}%
. This is an interesting observation that deserves further study.

In this paper we give an analytical proof of those numerical relations and
show that they are particular cases of a general property of the Hamiltonian
used. To this end we first derive complete analytical expressions for the
s.p. energies (these will be defined more precisely in Sec. 2), and then
derive the s.p. energy splittings in a major shell using the same
Hamiltonian as in Ref. \cite{FSZ}. We consider the problem of how to obtain
the s.p. energy splittings in a major shell with such an interaction which
will preserve the SU(3) results

More specifically, we use the Hamiltonian 
\begin{equation}
H=\frac{\hbar \omega _0}2\sum_i\left( {\bf x}_i^2+\mbox{\boldmath $\pi $}%
_i^2\right) -\frac \chi 2\sum_{ij}Q\left( i\right) \cdot (Qj)  \label{eq1}
\end{equation}
where 
\begin{equation}
Q\left( i\right) \cdot Q(j)=\sum_\mu Q_\mu \left( i\right) Q_\mu ^{\dagger
}(j)  \label{eq2}
\end{equation}

\begin{equation}
Q_\mu \left( i\right) =b^2x_i^2\,Y_2^\mu \left( \widehat{x}_i\right)
\label{eq3}
\end{equation}
with dimensionless coordinate and momenta ${\bf x}_i={\bf r}_i/b,%
\mbox{\boldmath $\pi $}_i={\bf p}_ib/\hbar $, and the harmonic oscillator
(H.O.) length scale $b=\sqrt{\hbar /m\omega _0}$.

\section{Definition of single particle energies and energy splittings}

To parallel the shell model calculations of Ref. \cite{FSZ}, we consider the
effect of the residual Q-Q interaction when we have a closed core and put a
valence nucleon in different orbital $(N\ell m)$ states of a major shell $%
(N) $ out of the closed core. In other words, we look for the s.p. energy
splitting between $N\ell $ and $N\ell ^{\prime }$ levels caused by the Q-Q
interaction. To this end we write the s.p. energy of the $N\ell $ level as 
\begin{equation}
E_{N\ell }=\hbar \omega _0\left( N+\frac 32\right) +E_{N\ell }^{Q1}+E_{N\ell
}^{Q2}  \label{eq4}
\end{equation}
where $E_{N\ell }^{Q1}$ comes from the diagonal part of the Q-Q interaction $%
\left( Q\left( i\right) \cdot Q\left( i\right) \right) $ 
\begin{equation}
E_{N\ell }^{Q1}=-\frac \chi 2<N\ell |Q\cdot Q|N\ell >  \label{eq5}
\end{equation}
or 
\begin{equation}
E_{N\ell }^{Q1}=-4\overline{\chi }<N\ell |x^4|N\ell >,\ \ \overline{\chi }%
=\chi \frac{5b^4}{32\pi }  \label{eq6}
\end{equation}
while $E_{N\ell }^{Q2}$ comes from the interaction of the valence particle
with the core. The direct term of the 2-body interaction $\left(
Q(i).Q(j),i\neq j\right) $ is proportional to the quadrupole moment of the
core, hence it is zero for a closed core. Consequently $E_{N\ell }^{Q2}$ is
given by the exchange term 
\begin{equation}
E_{N\ell }^{Q2}=\chi \sum_{N_c\ell _cm_c\mu }\left| <N\ \ell \ m\ |Q_\mu
|N_c\ \ell _c\ m_c>\right| ^2  \label{eq7}
\end{equation}

Clearly the s.p. energy defined in Eq. (\ref{eq4}) corresponds to the energy
difference of the systems with $A+1$ and $A$ (core) nucleons, calculated as
the expectation values of the Hamiltonian (\ref{eq1}) in the ground state of
the unperturbed $\left( \chi =0\right) $ harmonic oscillator Hamiltonian,
i.e., in the simplest, lower order harmonic oscillator shell model wave
function. We adopt this definition to follow the numerical work described in
Ref. \cite{FSZ}, although this is not the most general definition of s.p.
energy that may be considered with the Hamiltonian (\ref{eq1}).

Using the explicit expression for H.O. wave functions in terms of Laguerre
Polynomials we obtain the following result for $E_{N\ell }^{Q1}$ 
\begin{eqnarray}
E_{N\ell }^{Q1} &=&-4\overline{\chi }\left\{ n(n-1)+4n\left( n+\ell +\frac 32%
\right) \right.  \nonumber \\
&&\ \ +\left. \left( n+\ell +\frac 32\right) \left( n+\ell +\frac 52\right)
\right\}  \nonumber \\
&=&\overline{\chi }\,\left[ 2\ell \left( \ell +1\right) -3\left(
2N^2+6N+5\right) \right]  \label{eq8}
\end{eqnarray}
with $n$ the principal quantum number $\left( n=(N-\ell )/2\right) $(see for
instance \cite{int}).

To evaluate Eq. (\ref{eq7}) we have to keep in mind that according to the
selection rules for H.O. wave functions only matrix elements with $N_c=N-2$
will contribute to the sum over $N_c$. This is clear since the $Q_\mu $
operator only connects the states with $N$\ to those with $N^{\prime
}=N,N\pm 2$, and of these possible $N^{\prime }$ values only $N^{\prime
}=N_c=N-2$ belongs to the set of core levels. Carrying out the sum over $m_c$
and $\mu $ in Eq. (\ref{eq7}) we get 
\begin{equation}
E_{N\ell }^{Q2}=8\overline{\chi }\sum_{\ell _c}\left( <N\ \ell \ |x^2|N-2\
\ell _c>\right) ^2\left( 2\ell _c+1\right) \left( 
\begin{array}{ccc}
\ell & 2 & \ell _c \\ 
0 & 0 & 0
\end{array}
\right) ^2  \label{eq9}
\end{equation}

For H.O. wave functions the matrix element in Eq. (\ref{eq9}) is given by 
\begin{eqnarray}
<N\ \ell \ |x^2|N-2\ \ell ^{\prime }>= &&\delta _{\ell ^{\prime },\ell }%
\sqrt{n\left( n+\ell +\frac 12\right) }+\delta _{\ell ^{\prime },\ell +2}%
\sqrt{n\left( n-1\right) }  \nonumber \\
&&+\delta _{\ell ^{\prime },\ell -2}\sqrt{\left( n+\ell +\frac 12\right)
\left( n+\ell -\frac 12\right) }  \label{eq10}
\end{eqnarray}

Substitution of Eq. (\ref{eq10}) into Eq. (\ref{eq9}), together with the
following equalities that hold for $3-j$ symbols, 
\begin{equation}
\sum_{\ell ^{^{\prime }}}(2\ell ^{\prime }+1)\left( 
\begin{array}{ccc}
\ell & 2 & \ell ^{\prime } \\ 
0 & 0 & 0
\end{array}
\right) ^2=1  \label{eq11}
\end{equation}

\begin{equation}
\frac{(2\ell +3)}2\left( 
\begin{array}{ccc}
\ell & 2 & \ell \\ 
0 & 0 & 0
\end{array}
\right) ^2+(2\ell -3)\left( 
\begin{array}{ccc}
\ell & 2 & \ell -2 \\ 
0 & 0 & 0
\end{array}
\right) ^2=\frac \ell {2\ell +1}  \label{eq12}
\end{equation}

\begin{equation}
(2\ell -1)(2\ell -3)\left( 
\begin{array}{ccc}
\ell & 2 & \ell -2 \\ 
0 & 0 & 0
\end{array}
\right) ^2=\frac 32\frac{\ell (\ell -1)}{(2\ell +1)}  \label{eq13}
\end{equation}
leads to the following result for the {\it valence core interaction} 
\begin{eqnarray}
E_{N\ell }^{Q2} &=&8\overline{\chi }\left\{ n(n-1)+n\ \ell +\frac 38\ell
(\ell -1)\right\}  \nonumber \\
\ &=&\overline{\chi }\left[ \ell \left( \ell +1\right) +2N\left( N-2\right)
\right]  \label{eq14}
\end{eqnarray}

It is now a simple matter to compute the {\it energy splitting} $\Delta
_{N(\ell ,\ell ^{\prime })}$ between $N$-shell orbital partners $N\ell $ and 
$N\ell ^{\prime }$: 
\begin{equation}
\Delta _{N(\ell ,\ell ^{\prime })}\equiv \left( E_{N\ell }-E_{N\ell ^{\prime
}}\right) =3\overline{\chi }\left\{ \ell (\ell +1)-\ell ^{\prime }(\ell
^{\prime }+1)\right\}  \label{eq15}
\end{equation}
or for $\ell ^{\prime }=\ell -2$ 
\begin{equation}
\Delta _{N(\ell ,\ell ^{\prime }=\ell -2)}=6\overline{\chi }(2\ell -1)
\label{eq16}
\end{equation}
which agree with the SU(3) result. In particular in the $N=2$ and $N=3$
shells we recover the numerical results of Ref. \cite{FSZ}.

We can also show easily that one third of the energy splitting $\Delta
_{N(\ell ,\ell ^{\prime })}$ comes from the valence-core interaction and two
thirds from the diagonal term of Q-Q. Writing 
\begin{equation}
\Delta _{N(\ell ,\ell ^{\prime })}=\Delta _{N(\ell ,\ell ^{\prime
})}^{Q1}+\Delta _{N(\ell ,\ell ^{\prime })}^{Q2}  \label{eq17}
\end{equation}
with 
\begin{equation}
\Delta _{N(\ell ,\ell ^{\prime })}^{Q1}\equiv E_{N\ell }^{Q1}-E_{N\ell
^{\prime }}^{Q1}  \label{eq18}
\end{equation}
\begin{equation}
\Delta _{N(\ell ,\ell ^{\prime })}^{Q2}\equiv E_{N\ell }^{Q2}-E_{N\ell
^{\prime }}^{Q2}  \label{eq19}
\end{equation}
we obtain using Eqs. (\ref{eq8}) and (\ref{eq14}) 
\begin{equation}
\Delta _{N(\ell ,\ell ^{\prime })}^{Q1}=2\overline{\chi }\ \left[ \ell (\ell
+1)-\ell ^{\prime }(\ell ^{\prime }+1)\right]  \label{eq20}
\end{equation}
\begin{equation}
\Delta _{N(\ell ,\ell ^{\prime })}^{Q2}=\overline{\chi \ }\left[ \ell (\ell
+1)-\ell ^{\prime }(\ell ^{\prime }+1)\right]  \label{eq21}
\end{equation}
or 
\begin{equation}
\begin{array}{l}
\Delta _{N(\ell ,\ell ^{\prime })}^{Q1}=\frac 23\Delta _{N(\ell ,\ell
^{\prime })} \\ 
\Delta _{N(\ell ,\ell ^{\prime })}^{Q2}=\frac 13\Delta _{N(\ell ,\ell
^{\prime })}
\end{array}
\left\{ 
\begin{array}{c}
=4(2\ell -1)\overline{\chi }\text{ \quad for }\ell ^{\prime }=\ell -2 \\ 
=2(2\ell -1)\overline{\chi }\quad \text{ for }\ell ^{\prime }=\ell -2
\end{array}
\right\}  \label{eq22}
\end{equation}
which again particularize to the numerical results found in Ref. \cite{FSZ}
for the $N=2$ (1s-0d) and the $N=3$ (1p-0f) shells 
\begin{equation}
\begin{array}{ccc}
\Delta _{2(2,0)}^{Q1}=12\overline{\chi } & \ \ \  & \Delta _{3(3,1)}^{Q1}=20%
\overline{\chi } \\ 
\Delta _{2(2,0)}^{Q2}=6\overline{\chi } & \ \ \  & \Delta _{3(3,1)}^{Q2}=10%
\overline{\chi } \\ 
\Delta _{2(2,0)}=18\overline{\chi } & \ \ \  & \Delta _{3(3,1)}=30\overline{%
\chi }
\end{array}
\label{eq23}
\end{equation}

We note that although not explicitly mentioned, we have been considering the
valence particle in a shell close to the core. In principle we may as well
consider a valence particle in a higher $N$-shell, $N_v>N_c^{\max }+2$. In
that case the valence-core interaction is zero and only the $Q1$ splitting
remains, i.e., for these higher $N$-shells one gets a smaller splitting.

\section{Comparison with Elliott's Q$\cdot $Q interaction}

One may wonder how the one-third, two-thirds division of the s.p. energy
splitting between core-particle and diagonal contributions might be related
to the inclusion of the momentum-dependent parts of the Elliott quadrupole
operator $Q_\mu ^E$

\begin{equation}
Q_\mu ^E(i)=\frac 12b^2\left[ x_i^2\,Y_2^\mu \left( \hat x_i\right) +\pi
_i^2\,Y_2^\mu \left( \hat \pi _i\right) \right] \equiv \frac 12b^2\sqrt{%
\frac 5{4\pi }}\left( q_{r_i}^\mu +q_{p_i}^\mu \right)  \label{eq24}
\end{equation}
It is therefore instructive to study how the s.p. energy splitting is shared
by the position, momentum, and crossed terms of the Elliott $Q\cdot Q$
interaction

\begin{equation}
H_{QQ}^E=-\frac \chi 2\sum_{ij}Q^E\left( i\right) \cdot Q^E\left( j\right)
\label{eq25}
\end{equation}
We stress that this interaction is formally identical to that in Eq. (\ref
{eq1}). The only difference comes from the replacement of the position
quadrupole operator $Q$ by the $Q^E$ operator. The latter is a sum of the
dimensionless position and momentum quadrupole operators

\begin{equation}
q_r^\mu =\sqrt{\frac{4\pi }5}x_i^2\,Y_2^\mu \left( \hat x_i\right)
\label{eq26}
\end{equation}

\begin{equation}
q_p^\mu =\sqrt{\frac{4\pi }5}\pi _i^2\,Y_2^\mu \left( \hat \pi _i\right)
\label{eq27}
\end{equation}
We recall that an important property of Elliott's quadrupole operator is
that it has zero matrix elements between different $N-$shell states because
the $\Delta N=2$ matrix elements of $q_p$ exactly cancel those of $q_r$. As
can be seen from Table 1 the reduced matrix elements of $Q_\mu ^E$ are

\begin{eqnarray}
\left\langle N^{\prime }\ell ^{\prime }\left\| Q_\mu ^E\right\| N\ell
\right\rangle =\delta _{NN^{\prime }}\sqrt{\frac 5{4\pi }}b^2 &&\left\{
-\delta _{\ell ,\ell ^{\prime }}\left( N+\frac 32\right) \sqrt{\frac{\ell
\left( \ell +1\right) \left( 2\ell +1\right) }{\left( 2\ell -1\right) \left(
2\ell +3\right) }}\right.  \nonumber \\
&&\ +\delta _{\ell ^{\prime },\ell -2}\sqrt{\frac 32\frac{\left( N-\ell
+2\right) \left( N+\ell +1\right) \ell \left( \ell -1\right) }{2\ell -1}} 
\nonumber \\
&&\ \left. +\delta _{\ell ^{\prime },\ell +2}\sqrt{\frac 32\frac{\left(
N-\ell \right) \left( N+\ell +3\right) \left( \ell +1\right) \left( \ell
+2\right) }{2\ell +3}}\right\}  \label{eq28}
\end{eqnarray}
Therefore, with Elliott's interaction there is no particle-core interaction
and the s.p. energy defined in the previous section [Eq. (\ref{eq4})], only
gets a contribution from the diagonal term defined in Eq. (\ref{eq5}) with $%
Q $ replaced by $Q^E$. This diagonal contribution can be easily calculated
from Eq. (\ref{eq28}) and it is found to be

\begin{equation}
E_{N\ell }^{QE}=\bar \chi \left[ 3\ell \left( \ell +1\right) -4N\left(
N+3\right) \right]  \label{eq29}
\end{equation}
Hence, when $Q$ is replaced by $Q^E$ in Eq. (\ref{eq1}), the sum of the
one-body $\left( E_{N\ell }^{Q1}\right) $ and two-body $\left( E_{N\ell
}^{Q2}\right) $ contributions to the s.p. energy [Eq. (\ref{eq4})] is
replaced by the one-body contribution $E_{N\ell }^{QE}$. The resulting value
of the s.p. energy differs only in the $N-$dependence and therefore one is
left with identical s.p. energy splittings:

\begin{equation}
\Delta _{N\left( \ell ,\ell ^{\prime }\right) }^E=E_{N\ell }^{QE}-E_{N\ell
^{\prime }}^{QE}=3\bar \chi \left[ \ell \left( \ell +1\right) -\ell ^{\prime
}\left( \ell ^{\prime }+1\right) \right] \equiv \Delta _{N\left( \ell ,\ell
^{\prime }\right) }  \label{eq30}
\end{equation}
We also note that this s.p. energy splitting can be decomposed in three
contributions,

\begin{equation}
\Delta _{N\left( \ell ,\ell ^{\prime }\right) }^E=\Delta _{N\left( \ell
,\ell ^{\prime }\right) }^{E_r}+\Delta _{N\left( \ell ,\ell ^{\prime
}\right) }^{E_p}+\Delta _{N\left( \ell ,\ell ^{\prime }\right) }^{E_{rp}}
\label{eq31}
\end{equation}
one coming from the $q_r\cdot q_{r\text{ }}$interaction $\left( \Delta
^{E_r}\right) $, one coming from the $q_p\cdot q_{p\text{ }}$interaction $%
\left( \Delta ^{E_p}\right) $, and one coming from the crossed $q_r\cdot
q_p+ $ $q_p\cdot q_{r\text{ }}$ interaction $\left( \Delta ^{E_{rp}}\right) $%
. Using Table 1, it is straightforward to check that 
\begin{equation}
E_{N\ell }^{E_r}\equiv -\bar \chi \left\langle N\ell \left| q_r\cdot
q_r\right| N\ell \right\rangle =\bar \chi \left[ \frac 12\ell \left( \ell
+1\right) -\frac 34\left( 2N^2+6N+5\right) \right]  \label{eq32}
\end{equation}

\begin{equation}
E_{N\ell }^{Ep}\equiv -\bar \chi \left\langle N\ell \left| q_p\cdot
q_p\right| N\ell \right\rangle =E_{N\ell }^{E_r}  \label{eq33}
\end{equation}

\begin{equation}
E_{N\ell }^{E_{rp}}\equiv -\bar \chi \left\langle N\ell \left| q_r\cdot
q_p+q_p\cdot q_r\right| N\ell \right\rangle =\bar \chi \left[ 2\ell \left(
\ell +1\right) -\left( N^2+3N-\frac{15}2\right) \right]  \label{eq34}
\end{equation}
Therefore the SU(3) s.p. energy splitting $3\bar \chi \left[ \ell \left(
\ell +1\right) -\ell ^{\prime }\left( \ell ^{\prime }+1\right) \right] $ is
shared as one-sixth, one-sixth, and two-thirds by the position, momentum,
and crossed terms, respectively.

\[
\Delta _{N\left( \ell ,\ell ^{\prime }\right) }^{E_r}=\Delta _{N\left( \ell
,\ell ^{\prime }\right) }^{E_p}=\frac 16\Delta _{N\left( \ell ,\ell ^{\prime
}\right) } 
\]

\[
\Delta _{N\left( \ell ,\ell ^{\prime }\right) }^{E_{rp}}=\frac 23\Delta
_{N\left( \ell ,\ell ^{\prime }\right) } 
\]

\section{A note on spin-orbit interaction}

Since {\it spin-orbit coupling} is very important in nuclei it is also
useful to examine how the above results are affected when the one body
spin-orbit interaction $\left( V_{SO}=-2\chi _{SO}\;\mbox{\boldmath $\ell $}.%
{\bf s}\right) $, is added to the Hamiltonian. Different cases have to be
distinguished. The simplest cases are the s-d shell and f-p shell nuclei
where the closed cores consist of closed $N$-shells (N = Z = 8 or N = Z =
20). In these cases the results given in the previous section still hold for 
$N\ell $ and $N\ell ^{\prime }$ orbitals. The only difference is that one
may in addition consider the energy splitting between different $\ell j$ and 
$\ell ^{\prime }j^{\prime }$ subshells. Since the energy splitting between $%
\left( j=\ell -\frac 12\right) $ and $\left( j=\ell +\frac 12\right) $
partners is not perturbed by the Q-Q interaction, the total $N\ \ell \ j$- $%
N\ \ell ^{\prime }\ j^{\prime }$ splittings are for $\ell ^{\prime }=\ell -2$%
: 
\begin{equation}
\Delta _{N(\ell j,\ell ^{\prime }j^{\prime })}=6\overline{\chi }(2\ell
-1)+\chi _{S0}\ \left\{ 
\begin{array}{l}
\mp 2\text{ \quad for }j=\ell \pm \frac 12\text{ and }j^{\prime }=j-2 \\ 
-(2\ell -1)\text{ \quad for }j^{\prime }=j-3\ \left( j=\ell +\frac 12\right) 
\\ 
+(2\ell -1)\quad \text{ for }j^{\prime }=j-1\ \left( j=\ell -\frac 12\right) 
\end{array}
\right\}   \label{eq35}
\end{equation}
i.e., the spin-orbit in general destroys the SU(3) limit result, but for
some of the $\ell j$-$\ell ^{\prime }j^{\prime }$ splittings the
proportionality to $(2\ell -1)$ remains.

On the other hand when the closed core corresponds to magic numbers with N
or Z = 28, 50, etc., where only one of the $N\ell j$ subshells of the
valence shell $N$ (the $\ell j$ subshell with $n=0,\ell =N,j=\ell +\frac 12$%
) is closed, one has to take into account in addition the Q-Q interaction of
the valence particle with this subshell, i.e. when N or Z = 28, 50, 82, 126
the 2-body interaction contains also the term

\begin{equation}
\delta E_{N\ell j}^{Q2}=\chi \sum_{m_{jc}\mu }\left| <N\ell jm|Q_\mu |N\ell
_cj_cm_{jc}>\right| ^2  \label{eq36}
\end{equation}
with $\ell _c=N$ and $j_c=\ell _c+\frac 12=N+\frac 12$.

This contribution gives an extra term to the energy of the valence subshells
with quantum numbers 
\begin{equation}
(a)\ \ \ell =\ell _c=N,\ j=j_c-1=N-\frac 12  \label{eq37}
\end{equation}
and 
\begin{equation}
(b)\ \ \ell =\ell _c-2=N-2;\ j=\ell _c-2+\frac 12=N-\frac 32  \label{eq38}
\end{equation}
the value of this extra term is 
\begin{equation}
(a)\ \ \delta E_{N\ell j}^{Q2}=6\overline{\chi }\frac{(2N+3)(N+1)}{%
(2N+1)(2N-1)}\rightarrow 3\overline{\chi }\text{ for }N\text{ large}
\label{eq39}
\end{equation}
\begin{equation}
(b)\ \ \delta E_{N\ell j}^{Q2}=24\overline{\chi }\frac{N(N+1)}{2N-1}%
\rightarrow 12\overline{\chi }N\text{ for }N\text{ large}  \label{eq40}
\end{equation}

Thus, this extra contribution may also spoil the SU(3) limit [Eq.(\ref{eq15}%
)] for some of the s.p. energy splittings.

\section{Conclusions}

In summary, quite generally, the s.p. energy splitting in a major shell $N$
preserving the SU(3) result

\begin{eqnarray}
\Delta _{N(\ell ,\ell ^{\prime })} &\equiv &E_{N\ell }-E_{N\ell ^{\prime
}}=\Delta _{N(\ell ,\ell ^{\prime })}^{Q1}+\Delta _{N(\ell ,\ell ^{\prime
})}^{Q2}=\Delta _{N(\ell ,\ell ^{\prime })}^E=  \nonumber \\
\ &=&3\overline{\chi }\left[ <N\ell |{\bf L}^2|N\ell >-<N\ell ^{\prime }|%
{\bf L}^2|N\ell ^{\prime }>\right]  \label{eq41}
\end{eqnarray}
has been obtained with a Q-Q interaction in coordinate space that allows for
interactions between $\Delta N=2$ shells. As a general rule, the
valence-core interaction ---proportional to the matrix elements of the $Q$%
-operator between $\Delta N=2$ shells--- generates one third of the energy
splitting in Eq.(\ref{eq41}), while the other two thirds are due to the
diagonal (one-body) part of $Q\cdot Q$. This result is surprising since
traditionally the SU(3) limit of Elliott's model was derived restricting the
action of the Q-Q interaction to a single major shell (eliminating the $%
\Delta N=2$ matrix elements of the $Q$-operator).

We have also explicitly shown that with Elliott quadrupole operator the same
s.p. energy splitting is obtained by adding up the contributions from the
position, momentum, and crossed position-momentum terms of the Elliott
quadrupole-quadrupole interaction. This comparison is important because it
illustrates how in this instance, the same effect can be obtained by taking
into account (2-body) particle-hole interactions or (1-body) momentum
dependent interactions. The fact that the s.p. energy splittings are equal
while the s.p. energies are not, also serves to illustrate how the two types
of interactions can be equivalent in some aspects while differing in others.

Our conclusion here generalizes and reinforces the observation made in Ref. 
\cite{FSZ} for the $N=2,3$ shells and suggests new ways for further
applications of the model Hamiltonian (\ref{eq1}). Since applications of
this model Hamiltonian in the past have proven to be very fruitful it is
worth to explore it in a new direction. A clear practical application
involves deriving the $Q\cdot Q$ interaction from the realistic interaction.
One now does not have to justify the momentum terms.

We have also found that although the spin-orbit coupling tends to spoil the
SU(3) limit in Eq. (\ref{eq41}), the proportionality to $(2\ell +3)$ of the
s.p. energy splitting between $(\ell ,j)$ and $(\ell +2,j^{\prime })$
subshells is still maintained in some cases.

\acknowledgments 

This work was supported by DGICYT (Spain) under contract number PB95/0123
and by a Department of Energy grant DE-FG05-95ER-40940.

\newpage

\begin{table}[tbp]
{\bf Table 1. }Reduced matrix elements of quadrupole operators in $r-$ and $%
p-$spaces.
\par
Note that $\left\langle N\ell ^{\prime }\left\| q_p\right\| N\ell
\right\rangle =\left\langle N\ell ^{\prime }\left\| q_r\right\| N\ell
\right\rangle $ , $\left\langle \left( N\pm 2\right) \ell ^{\prime }\left\|
q_p\right\| N\ell \right\rangle =-\left\langle \left( N\pm 2\right) \ell
^{\prime }\left\| q_r\right\| N\ell \right\rangle $.
\par
\begin{tabular}{cccc}
& $\left\langle N\ell ^{\prime }\left\| q_p\right\| N\ell \right\rangle $ & $%
\left\langle \left( N-2\right) \ell ^{\prime }\left\| q_p\right\| N\ell
\right\rangle $ & $\left\langle \left( N+2\right) \ell ^{\prime }\left\|
q_p\right\| N\ell \right\rangle $ \\ \hline
$\ell ^{\prime }=\ell $ & $-\left( N+\frac 32\right) \sqrt{\frac{\ell \left(
\ell +1\right) \left( 2\ell +1\right) }{\left( 2\ell -1\right) \left( 2\ell
+3\right) }}$ & $\sqrt{\frac{n\left( n+\ell +1/2\right) \ell \left( \ell
+1\right) \left( 2\ell +1\right) }{\left( 2\ell -1\right) \left( 2\ell
+3\right) }}$ & $\sqrt{\frac{\left( n+1\right) \left( n+\ell +3/2\right)
\ell \left( \ell +1\right) \left( 2\ell +1\right) }{\left( 2\ell -1\right)
\left( 2\ell +3\right) }}$ \\ 
$\ell ^{\prime }=\ell -2$ & $\sqrt{6\frac{\left( n+1\right) \left( n+\ell
+1/2\right) \ell \left( \ell -1\right) }{2\ell -1}}$ & $-\sqrt{\frac 32\frac{%
\left( n+\ell +1/2\right) \left( n+\ell -1/2\right) \ell \left( \ell
-1\right) }{2\ell -1}}$ & $-\sqrt{\frac 32\frac{\left( n+1\right) \left(
n+2\right) \ell \left( \ell -1\right) }{2\ell -1}}$ \\ 
$\ell ^{\prime }=\ell +2$ & $\sqrt{6\frac{n\left( n+\ell +3/2\right) \left(
\ell +1\right) \left( \ell +2\right) }{2\ell +3}}$ & $-\sqrt{\frac 32\frac{%
n\left( n-1\right) \left( \ell +1\right) \left( \ell +2\right) }{2\ell +3}}$
& $-\sqrt{\frac 32\frac{\left( n+\ell +5/2\right) \left( n+\ell +3/2\right)
\left( \ell +1\right) \left( \ell +2\right) }{2\ell +3}}$%
\end{tabular}
\end{table}

\end{document}